\documentclass{article}

\usepackage[english]{babel}
\usepackage{amsmath,amssymb,amsfonts,amsthm}
\usepackage{hyperref}
\usepackage{graphicx}
\usepackage{url}
\usepackage{acronym}
\usepackage{subcaption}
\usepackage{colortbl}
\usepackage{xfrac}
\usepackage[inline]{enumitem}
\usepackage[numbers]{natbib}
\usepackage{textcomp}
\usepackage[htt]{hyphenat}

 \newtheorem{example}{Example}

\definecolor{Gray}{gray}{0.80}
\definecolor{nfyellow}{rgb}{1.00, 0.75, 0.0}

\newcommand{\quotes}[1]{``.1"}

\begin{document}

\title{Response to Moffat's Comment on ``Towards Meaningful Statements in IR Evaluation: Mapping Evaluation Measures to Interval Scales''}

\author{Marco Ferrante\thanks{Department of Mathematics ``Tullio Levi-Civita'', University of Padua, Italy (email: ferrante@math.unipd.it)}, Nicola Ferro\thanks{Department of Information Engineering, University of Padua, Italy (email: ferro@dei.unipd.it)}, Norbert Fuhr\thanks{Faculty of Engineering, University of Duisburg-Essen, Germany (email: norbert.fuhr@uni-due.de)}}

\maketitle

\begin{abstract}
    \citeauthor{Moffat2022} recently commented on our previous work published in the IEEE Access journal. Our work focused on how laying the foundations of our evaluation methodology into the representational theory of measurement can improve our knowledge and understanding of the evaluation measures we daily use in \acf{IR} and how it can shed light on the different types of scales adopted by our evaluation measures; we also provided evidence, through extensive experimentation, on the impact and effect of the different types of scales on the subsequent statistical analyses, as well as on the impact of departing from their assumptions. Moreover, we investigated, for the first time in \ac{IR}, the concept of \emph{meaningfulness}, intended as a specific sort of invariance of the experimental statements and inferences you draw, and proposed it as a way to ensure more valid and generalizabile results.
    
    \citeauthor{Moffat2022}'s comments build on: (i) misconceptions about the representational theory of measurement, such as what an interval scale actually is and what axioms it has to comply with; (ii) they totally miss the central concept of \emph{meaningfulness}. Therefore, we reply to \citeauthor{Moffat2022}'s comments by properly framing them in the representational theory of measurement and in the concept of \emph{meaningfulness}. 

    All in all, we can only reiterate what we said several times: the goal of this research line is to theoretically ground our evaluation methodology -- and \ac{IR} is a field where it is extremely challenging to perform any theoretical advances -- in order to aim for more robust and generalizable inferences -- something we currently lack in the field. Possibly there are other and better ways to achieve this objective and these proposals could emerge from an open discussion in the field and from the work of others. On the other hand, reducing everything to a contrast on what is (or pretend to be) an interval scale or whether all or none evaluation measures are interval scales may be more a barrier from than a help in progressing towards this goal.
\end{abstract}

\section{Introduction}

We write this response in reply to \citeauthor{Moffat2022}'s comment \citep{Moffat2022} on our paper ``Towards Meaningful Statements in IR Evaluation: Mapping Evaluation Measures to Interval Scales'' \citep{FerranteEtAl2021c}.

We hope that this response will help to clarify some misunderstandings about the \emph{representational theory of measurement}~\citep{KrantzEtAl1971,LuceEtAl1990,SuppesEtAl1989}, in general, and the notion of \emph{meaningfulness}~\cite{Narens2002} in particular -- misconceptions central to \citeauthor{Moffat2022}'s comment and leading to somehow fallacious arguments. We also hope that this response will contribute to a \emph{constructive and informed discussion} in the field about these themes, which are often faced just as a contrast of opinions.

\acf{IR} is an highly experimental field by nature and by necessity, where we suffer from a lack of \emph{generalizability} of our results and, consequently, the (almost) impossibility of \emph{predicting} the performance of our systems. There is already some agreement on the need for ``a better understanding of the assumptions and user perceptions underlying different metrics, as a basis for judging about the differences between methods'' and on the fact that ``the assumptions underlying our algorithms, evaluation methods, datasets, tasks, and measures should be identified and explicitly formulated. Furthermore, we need strategies for determining how much we are departing from these assumptions in new cases and how much this impacts on system performance''~\citep{FerroEtAl2018}.  In this respect, our work adheres to the representational theory of measurement and proceeds step-by-step along its lines, in order to understand and check assumptions and deviations as well as their implications and consequences. 

We welcome very much the comment by \citeauthor{Moffat2022}, as an exemplar way: (i) to open the discussion to the community; (ii) to keep a written record of the arguments, record which, today, helps researchers in taking informed decisions and, in the future, will remain useful to motivate adopted approaches or to further revise them; and, (iii) to have a transparent and frank exchanges of views on a somewhat heartfelt topic. On the other hand, we would very much prefer all of this not to be framed as a contrast between, in \citeauthor{Moffat2022}'s words, ``a bleak picture of past decades of IR evaluation'' and ``a more optimistic view of IR evaluation and IR measurement'', because this rhetoric is apt to fuel oppositions, leaving everything unchanged, rather than to tackle the real issue, perhaps to favor the formulation of alternative and better solutions, and to make overall progress.

The paper is organized as follows: Section~\ref{sec:arguments} summarizes the main arguments in~\citeauthor{Moffat2022}'s comment; Section~\ref{sec:preliminaries} introduces preliminary notions on the representation theory of measurement and on \emph{meaningfulness} in order to set a proper stage for the discussion; Section~\ref{sec:response} provides our response to \citeauthor{Moffat2022}'s main arguments; finally, Section~\ref{sec:conclusions} reports some concluding remarks.

\section{Main Arguments by Moffat}
\label{sec:arguments}

As \citeauthor{Moffat2022}'s comment does not refer to any specific part of our paper to which we could answer, we will start from his text instead, considering his main arguments by directly quoting them. We also limit ourselves to the very major arguments, avoiding to enter in too low level or too technical details, because this would make our response too long, less clear, and of little use to a researcher who wants to form their opinion or to contribute new ideas.

\begin{description}
    \item[\textbf{MA1}] ``If you do know where the numbers came from and why they have the values that they do, and are confident that those values can be justified in reference to the real world attribute that the mapping is designed to represent, then those numbers may be used in your analysis and interpretation of that real world equivalence''~\citep[p.~105570]{Moffat2022}
        \begin{description}
            \item[\emph{MA1.1}] ``While care needs to be exercised when choosing the metric that best fits the user experience for any particular IR application [...] once that match has been decided, the values calculated by the effectiveness metric may be used as simple numbers ``that don't remember where they came from''~\citep{Lord1953}; that is, without regard to their origins in a categorical-scale SERP dataset''~\citep[p.~105576]{Moffat2022}
        \end{description}
        
    \item[\textbf{MA2}] ``Via a sequence of examples we have presented our view that all IR effectiveness metrics can be considered to be interval scale measurements, provided only that the mapping from SERP categories to numeric scores has a real-world basis (an external validity) and can be motivated as corresponding to the underlying usefulness of each SERP, as experienced by an identified cohort of users as they carry out some identified search task''~\citep[p.~105576]{Moffat2022}
        \begin{description}
            \item[\emph{MA2.1}] ``There can be no ambiguity: RR is an interval scale measurement for those users''~\citep[p.~105573]{Moffat2022}
            \item[\emph{MA2.2}]  ``Any argument that RR -- or any other metric -- is an unsuitable categorical to numeric mapping for measuring IR system effectiveness for some cohort of users or some type of search task must be justified based on rhetoric about user perceptions of SERP usefulness, or on observational data that measures SERP usefulness via some agreed surrogate. Arguments against IR effectiveness metrics cannot be based solely upon statements about the non-uniformity of the intervals between the available measurement points''~\citep[pp.~105573--105574]{Moffat2022}
            \item[\emph{MA2.3}] ``We argue that most current IR metrics are well-founded, and, moreover, that those metrics are more meaningful in their current form than in the proposed ``intervalized'' versions''~\citep[p.~105564]{Moffat2022}
        \end{description}

    \item[\textbf{MA3}] ``We believe that intervalization should regarded with scepticism. There is no requirement in Steven's typology that interval scales be restricted to uniform distances between the available measurement points; the requirement is simply that the ratio between pairs of intervals be indicative of the corresponding difference in the underlying attribute''~\citep[p.~105574]{Moffat2022}
        \begin{description}           
            \item[\emph{MA3.1}] ``We have argued that the proposed intervalization of current IR effectiveness metrics is neither required nor helpful. If the raw metric value is indeed a defensible measurement of SERP usefulness and corresponds to the user's experience when they are presented with a member of that SERP category, then equi-intervalizing those measurements via a different categorical to numeric mapping must of necessity distort and alter any findings that arise, and thus risk masking what would otherwise be valid conclusions. And if the raw metric is not a defensible measurement of SERP usefulness for the search task at hand, then equi-intervalizing its scores is unlikely to improve the situation''~\citep[p.~105576]{Moffat2022}
            \item[\emph{MA3.2}] ``Moreover, altering the categorical to numeric mapping used to assign score to SERPs changes the relativities being measured, and thus affects the outcome of any subsequent arithmetic''~\citep[p.~105574]{Moffat2022}
        \end{description}
\end{description}

\section{Preliminaries on Measurement and Meaningfulness}
\label{sec:preliminaries}

Although we spent quite some effort to explain the following concepts in our original article~\citep{FerranteEtAl2021c}, as well as in previous and more formal work~\citep{FerranteEtAl2018b}, providing detailed discussion, references, and examples, we summarize here the main concepts needed to articulate our response. In doing so, we also opt for quoting as many passages as possible directly from the foundational works in the area of measurement, in order to plainly report third party research, without any additional interpretation on our side.

An introductory presentation of the concepts described below can be found in general textbooks about measurement, such as~\citet{Hand2010}, or textbooks about software metrics, such as~\citet{FentonBieman2014}. For a historical overview of the evolution of the theory of measurement and its concepts, you can refer to~\citet{Diez1997a,Diez1997b}.

\subsection{Measurement and Scales}
\label{subsec:scales}

``When measuring some attribute of a class of objects or events, we associate numbers [...] with objects in such a way that the properties of the attribute are faithfully represented as numerical properties''~\citep[p.~1]{KrantzEtAl1971}.

Suppose we aim at measuring the \emph{length} of rods. In the real world, we can compare rods to determine which one is longer, i.e. we have a comparison relation $\succ$ among rods; we can also concatenate rods together, i.e. we have a concatenation operation $\circ$ among rods.

``A \emph{relational structure} is a set with one or more relations on that set''~\citep[p.~8]{KrantzEtAl1971}.

So, in our example, we have a set of rods $A$, a binary relation $a \succ b$ to compare them, and a ternary relation $c = a \circ b$ to concatenate them. Overall, $\langle A, \succ, \circ \rangle$ is an empirical relational structure over the set of rods; it is called empirical because it exists in the real world.

``The numerical assignment $\phi$ is a \emph{homomorphism} in the sense that it sends $A$ into $\mathbb{R^+}$, $\succ$ into $>$, and $\circ$ into $+$ in such a way that $>$ preserves the properties of $\succ$ and $+$ the properties of $\circ$''~\citep[p.~8]{KrantzEtAl1971}, where the properties are~\citep[p.~4]{KrantzEtAl1971}:
\begin{enumerate}
    \item $a \succ b$ if and only if $\phi(a) > \phi(b)$;
    \item $\phi(a \circ b) = \phi(a) + \phi(b)$.
\end{enumerate}

These small excerpts from \citet{KrantzEtAl1971}, the first volume of the landmark three-volume series on the foundations and mathematical formalization of measurement, provide us with an intuitive understanding of the basic idea behind the measurement theory: we use numbers as proxies of attributes of real world objects, provided that the relations and operations among those numbers keep corresponding to the relations and operations among real world objects. 

This is formulated mathematically by saying that the assignment $\phi$, which is called a \emph{scale}, is, in general, a homomorphism. Note that the homomorphism is used because $\phi(a) = \phi(b)$ does not necessarily mean that the rods $a$ and $b$ are the same rod but just that they have the same length. However, if we consider the equivalence relation $\sim$ on $A$, then the empirical relational structure $\langle A/{\sim}, \succeq, \circ \rangle$ on the quotient set $A/{\sim}$ becomes an \emph{isomorphism} to the numerical relational structure $\langle \mathbb{R}^+, \geq, + \rangle$. This further underlines the \emph{bijective} nature of the correspondence between real world objects and numbers, as well as their relations and operations.

The theory of measurement moves a further step forward and asks and additional question: ``Given a set of rods, a comparison relation $\succ$, and a concatenation operation $\circ$, what assumptions concerning $\succ$ and $\circ$ are necessary and/or sufficient to construct a real-valued function $\phi$ that is order preserving and additive?''~\citep[p.~8]{KrantzEtAl1971}. Therefore, we have to seek for characteristics of the real world objects, expressed in the form of axioms, which guarantee the existence of a numerical scale $\phi$ with the desired properties. More specifically, ``a \emph{representation theorem} asserts that if a given relational structure satisfies certain axioms, then a homomorphism into a certain numerical relational structure can be constructed [...] Measurement can be regarded as the construction of homomorphisms (scales) from empirical relational structures of interest into numerical relational structures that are useful''~\citep[p.~9]{KrantzEtAl1971}.

However, the procedure for numerical assignment may look somewhat arbitrary or there may exist several alternatives which are equally good. For example, in the case of rods, which rod size is chosen as unit is an arbitrary matter, leading to equivalent scales. In other cases, it may be less immediate to see which choices are arbitrary and which are not. For example, when we count how many times the unit rod $u$ has to be concatenated for obtaining a given rod $a$, why do we record $n$ instead of $n^2$ or $\mathrm{e}^n$?

The notion of \emph{permissible transformation} serves exactly the purpose of answering this question: ``A transformation $\phi \rightarrow \phi^\prime$ is permissible if and only if $\phi$ and $\phi^\prime$ are both homomorphisms of $\langle A, R_1, \ldots, R_n \rangle$ into the \emph{same} numerical structure $\langle \mathbb{R}, S_1, \ldots, S_n \rangle$''~\citep[p.~12]{KrantzEtAl1971}. An \emph{uniqueness theorem} has the purpose of determining what this permissible transformation is and this is often not obvious at all.

For example, in the case of rod and length, \citet{Holder1901} developed the first proof that if $\phi$ is order preserving and additive, i.e. a homomorphism of $\langle A, \succ, \circ \rangle$ into  $\langle \mathbb{R}^+, >, + \rangle$, the same is true for $\alpha\phi$ when $\alpha > 0$; moreover, if $\phi^\prime$ is \emph{any} homomorphism of $\langle A, \succ, \circ \rangle$ into  $\langle \mathbb{R}^+, >, + \rangle$, then $\phi^\prime = \alpha\phi$ for some $\alpha > 0$.

The uniqueness theorem grasps the fact that even if we refer to different scales, e.g. meters or feet for length, they are actually equivalent from the standpoint of the permissible transformation and they are required to be so, otherwise the homomorphism with the real world $\langle A, \succ, \circ \rangle$ would be lost.

In this context, \citet{Stevens1946} defined the different types of scales, i.e. different types of $\phi$ which have to comply with specific axioms in order to guarantee desired properties. They can be briefly summarized as follows:
\begin{itemize}
    \item \textbf{Nominal scale}: it is used when entities of the real world can be placed into different classes or categories on the basis of their attribute under examination, without any notion of ordering among them. Any distinct numeric representation of the classes is an acceptable measure but there is no notion of magnitude associated with numbers. 
    
    Therefore, any arithmetic operation on the numeric representation has no meaning. As a consequence, the only allowable statistics is counting number of items in each class, that is mode and frequency.
    
    The class of permissible transformations is the set of all \emph{one-to-one mappings}, i.e. bijective functions: $\phi^\prime = \mathrm{f}(\phi)$, since they preserve the distinction among classes.

    \item \textbf{Ordinal scale}: it can be considered as a nominal scale where, in addition, there is a notion of ordering among the different classes or categories. Any distinct numeric representation which preserves the ordering is acceptable. Therefore, the magnitude of the numbers is used just to represent the ranking among classes.
    
    Addition, subtraction or other mathematical operations have no meaning. As a consequence, besides the statistics already allowed for nominal scales, median, quantiles, and percentiles are appropriate, since there is a notion of ordering.
    
    The class of permissible transformations is the set of all the \emph{monotonic increasing functions}, since they preserve the ordering: $\phi^\prime = \mathrm{f}(\phi)$.

    \item \textbf{Interval scale}: besides relying on ordered classes, it also captures information about the size of the intervals that separate the classes.  An interval scale preserves order, as an ordinal one, and differences among classes have meaning -- but not their ratio. 
    
    Addition and subtraction are acceptable operations but not multiplication and division. As a consequence, besides the statistics allowed for ordinal scales, mean and standard deviation are allowable since they depend just on sum and subtraction\footnote{Note that when we talk about admissible operations, we mean operations between items of the scale. So, for example, a mean involves summing items of the scale, e.g. temperature, and this is possible on an interval scale. The fact that a mean also requires a division by the number $N$ of items added together is not in contrast with saying that only addition and subtraction are allowed, since $N$ is not an item of the scale.}.
    
    The class of permissible transformations is the set of all the \emph{affine transformations}: $\phi^\prime = \alpha\phi + \beta, \, \alpha > 0$.

    Temperature is the typical example of an interval scale.

    \item \textbf{Ratio scale} it allows us to compute ratios among the different classes since classes, which are ordered. 
    
    All the arithmetic operations are allowed. As a consequence, besides the statistics allowed for interval scales, geometric and harmonic mean, as well as coefficient of variation, are allowable since they depend on multiplication and division.
    
    The class of permissible transformations is the set of all the \emph{linear transformations}: $\phi^\prime = \alpha\phi, \, \alpha > 0$.

    Length and mass are typical examples of ratio scales.
   
\end{itemize}

Let us focus on interval scales, which are the matter of discussion in our paper and in \citeauthor{Moffat2022}'s comment.

The important characteristics of interval scales is that they need to be based on \emph{equally spaced} objects in the real world which, in turn, leads to equi-spacing of the numerical mapping. 

\citeauthor{Stevens1946} does not explicitly mention the term ``equi-spacing'' in Table~1 at page~678, where he summarizes the scale types. However, he explicitly says this in the section where he explains what interval scales are: ``most psychological measurement aspires to create interval scales, and it sometimes succeeds. The problem usually is to devise operations for \emph{equalizing the units} of the scales''~\citep[p.~679]{Stevens1946}, providing also concrete examples like ``\emph{equal intervals} of temperature are scaled off by noting \emph{equal volumes}
of expansion''~\citep[p.~679]{Stevens1946}. Other sentences like ``the scale form remains invariant when a constant is added''~\citep[p.~679]{Stevens1946} or ``if the purpose of the scale is still served when its values are squared or cubed, it is not even an interval scale''~\citep[p.~680]{Stevens1946} implicitly assume an equally spaced scale. 

Let us now see how an interval scale is defined from a formal point of view. \citet[p.~56]{Rossi2014} explains ``How can we assign numbers (measures) to them [objects] in such a way that they properly express both the order and the distances? For doing so, we have to establish an \emph{equally spaced graduation}''. In particular, the equally spaced graduation is formulated in terms of a so-called \emph{solvability condition}, requiring that whenever we have two not equivalent intervals, it is always possible to find elements which correspond to each treat of the equally spaced graduation~\citet[p.~57]{Rossi2014}. 

All the properties required required to real world object to allow for the creation of an interval scale are determined by the definition of a \emph{difference structure}, i.e. specific type of empirical relational structure. Axiom 12.4 of Definition 3.12 (Difference Structure)~\citep[p.~59]{Rossi2014} expresses the \emph{solvability condition}:
\begin{equation*}
    \begin{aligned}
    \text{if } & \Delta_{ab} \succeq_d \Delta_{cd} \succeq_d \Delta_{aa} \\
     & \text{then there exist } d^\prime \in A \text{ and } d^{\prime\prime} \in A \\
     & \text{so that } \Delta_{ad^\prime} \sim_d \Delta_{cd} \sim_d \Delta_{d^{\prime\prime}b}
    \end{aligned}
\end{equation*}
where $\Delta_{ab}$ is the \emph{difference} in the real world, not among numbers, and $\succeq_d$ is a weak order among differences, again in the real world.

The definition of difference structure is then used in the \emph{Representation Theorem} 3.17 to demonstrate the existence of an interval scale~\citep[p.~62]{Rossi2014}:
\begin{equation*}
  \Delta_{ab} \succeq_d \Delta_{cd} \iff \phi(a) - \phi(b) \geq \phi(c) - \phi(d)
\end{equation*}
based on the equivalence${\iff}$between the notion of difference in the real world, consisting of equi-space objects, and the notion of different in the numerical system.

 Finally, the \emph{Uniqueness Theorem} 3.18 demonstrates that the affine transformation $\phi^\prime = \alpha\phi + \beta$, $\alpha>0$, is the permissible transformation for an interval scale~\citep[p.~62]{Rossi2014}. 

 A similar formalization is adopted even more extensively by \citet[pp.~136ff.]{KrantzEtAl1971}.

As a consequence of the above formal definition of interval scale, the ratio of differences among classes, i.e. the ratio of intervals, is allowed and invariant to an affine transformation~\citep[p.~10]{KrantzEtAl1971}: 
\begin{equation*}
\frac{\phi^\prime(a) - \phi^\prime(b)}{\phi^\prime(c) - \phi^\prime(d)} = \frac{[\alpha\phi(a) + \beta] - [\alpha\phi(b) + \beta]}{[\alpha\phi(c) + \beta] - [\alpha\phi(d) + \beta]} = \frac{\phi(a) - \phi(b)}{\phi(c) - \phi(d)}
\end{equation*}

\subsection{Meaningfulness}
\label{subsec:meaningfulness}

\emph{Meaningfulness} is a technical term in the theory of measurement, which relies on a precise mathematical formulation and serves the purpose of developing a whole theory around it. It should not be confused with ``meaningful'' in the everyday language sense, i.e. making sense or being credible.

For a rigourous and formal definition of \emph{meaningfulness}, please, refer to~\citet{Narens2002}. For the purpose of the present discussion, the intuitive definition by~\citet[pp.~99-100]{AdamsEtAl1965} should suffice: ``the criterion of appropriateness for a statement about a statistical operation is that the statement be \emph{empirically meaningful} in the sense that its truth or falsity must be invariant under permissible transformations of the underlying scale''.

Therefore, meaningfulness focuses on the \emph{invariance} of the statements we make and not on how much sense the make to us or how much true or false they are. A statement like ``A Chihuahua dog is three times taller than a Great Dane dog'' is false (possibly it does not make much sense either) and it stays false independently from whether we are using meters or feet, i.e. under a permissible linear transformation of the scale.

\emph{Meaningfulness} is not different from the notion of \emph{invariance} we have in geometry, when we ask that a shape remains the same independently from translation or rotation. We may wonder if \emph{meaningfulness} is asking too much or a too strong property to a scale. But, asking that the truth value of a statement remains the same under permissible transformations, which we have seen to be an intrinsic and indispensable property of a scale, is the same as asking that we draw the same inferences and conclusions independently from whether we are using meters or feet. All of this does not seem much stricter than asking, when we are looking at a cube, that, if we rotate it by 30 degrees, we still see a cube and not a sphere instead.

To clarify how \emph{meaningfulness} works, for example in the case of an interval scale, we report here an example taken from our previous paper~\cite{FerranteEtAl2021c}.

\begin{example}[Meaningfulness for an Interval Scale]
	The statement `Today the difference in temperature between Rome and Oslo is twice as high as it was one month ago'' is meaningful. Indeed,  if, on the Celsius scale, the temperature today in Rome is $20$ $^\circ$C and in Oslo is $10$ $^\circ$C while one month ago it was $12$ $^\circ$C and $7$ $^\circ$C, leading to $20 - 10 = 10$ which is twice as  $12 - 7 = 5$, on the Fahrenheit scale we would have $68 - 50 = 18$ which is twice as $53.6 - 44.6 = 9$.

	Suppose now that we have recorded two sets of temperatures from Paris and Rome: $T_P^C = [2\; 2\; 4\; 8\; 36]$ and $T_R^C = [1\; 2\; 4\; 15\; 34]$ in Celsius degrees and, the same, $T_P^F = [35.6\; 35.6\; 39.2\; 46.4\; 96.8]$ and $T_R^F = [33.8\; 35.6\; 39.2\; 59.0\; 93.2]$ in Fahrenheit degrees. 
	
	The statement ``The median temperature in Paris is the same as in Rome'' is meaningful, since $4 = 4$ in Celsius degrees and $39.2 = 39.2$ in Fahrenheit degrees; this is due to the fact that interval scales are also ordinal scales and quantiles are an allowable operation on ordinal scales. 
	
	The statement ``The mean temperature in Paris is less than in Rome'' is meaningful as well, since $10.4 < 11.2$ in Celsius degrees and $50.72 < 52.16$ in Fahrenheit degrees; this is due to the fact that addition and subtraction are allowable operations on an interval scale and, as a consequence, mean is allowable as well. 
 
	Finally, the statement ``The geometric mean of temperature in Paris is greater than in Rome'' is not meaningful, since $5.40 > 5.27$ in Celsius degrees and $46.74 < 48.17$ in Fahrenheit degrees; this is due to the fact that the geometric mean involves the multiplication and division of values, which is not a permitted operation on an interval scale. 
 
	\label{ex:interval-scale-meaningfulness}
\end{example}

Examples along this lines could be done for an ordinal scale showing that the median (or any other percentile) remains \emph{meaningful} for any monotonic increasing transformation but not the mean (or even the geometric mean).

\section{Response to Main Arguments by Moffat}
\label{sec:response}

\subsection{MA1: If you do know where the numbers came from$\ldots$}
\label{subsec:MA1}

We all agree that numbers should be chosen in such a way to reflect the attribute of a real world object and, in this respect, taking into account also the user experience does not make an exception.

However, at the same time, from the discussion in Section~\ref{subsec:scales}, it should be clear that operations among the numbers should keep corresponding to operations among real world objects and that numbers should keep their link with the attribute of the real world object.

More formally, the theory of measurement explicitly says that you have to create an \emph{homomorphism} between the real world and the numerical system in order to ensure that relations and operations are preserved. Even more, the theory of measurement makes a step further and states that among real world objects certain properties should hold in order to ensure both the existence (\emph{representation theorem}) of a mapping to numbers with desired properties and its uniqueness (\emph{uniqueness theorem}). This is, for example, the case of the \emph{difference structure}~\citep[Section 3.4, pp.~55ff.]{Rossi2014} and~\citep[Chapter 4, pp.~136ff.]{KrantzEtAl1971} which impose axioms among the real world objects that allow for constructing an interval scale.

\citeauthor{Moffat2022} reported the famous statement by~\citet{Lord1953} who, via the opinion of a statistician in his story, says: ``since the numbers don't remember where they came from, they always behave just the same way, regardless''. This statement is often used to support the use of whatever operation on numbers, regardless of any scale consideration and justified by practical usefulness. It is well known that \citeauthor{Lord1953}'s paper has been debated for decades and we reported the different viewpoints on it in our original paper~\citep{FerranteEtAl2021c}. 

In particular, \citet{ScholtenBorsboom2009}, referred also by~\citeauthor{Moffat2022}, show that \citeauthor{Lord1953}'s argument is broken and you cannot compute a mean on an ordinal scale just because ``numbers don't remember where they come from''. On the contrary, \citeauthor{ScholtenBorsboom2009} demonstrate, also formally in their appendix, that the reason why the means computed by the statistician in \citeauthor{Lord1953}'s paper can work is completely different and not related to \citeauthor{Lord1953}'s statement. Indeed, the question these means are answering is not about ``football players numbers'', as \citeauthor{Lord1953} assumes, but about the bias in the machines assigning those numbers to players and how they have been tampered with. In other terms, the means are about a different attribute of a different real world object (the machines and not the football players) and \citeauthor{ScholtenBorsboom2009} showed that on this new attribute it is possible to define a \emph{bisymmetrical structure}\footnote{Note that the \emph{bisymmetrical structure} used by \citet{ScholtenBorsboom2009} to define an interval scale relies on a \emph{solvability condition} (equi-spacing) like those used in the \emph{difference structures} mentioned in Section~\ref{subsec:scales}.} which, in turn, leads to an interval scale. Therefore, \citeauthor{ScholtenBorsboom2009} demonstrate that even the \citeauthor{Lord1953}'s argument, when properly conceptualized, leads to and supports \citeauthor{Stevens1946}'s definition of scales and the admissible operations on them.

Moreover, \citeauthor{Moffat2022}'s improperly uses the term \texttt{meaningfulness} in its everyday language sense instead of its scientific one and this leads to wrong conclusions. Sentences like ``that factual relationship makes the mapping's values \texttt{meaningful}, and hence interpretable in terms of the attribute from which the measurement was derived''~\citep[p.~105569]{Moffat2022} or ``the intervals between the salary points are \texttt{meaningful}; they represent salary differentials that must be paid in a competitive market, measured in dollars''~\citep[p.~105569]{Moffat2022} are wrong and, certainly, do not support \citeauthor{Moffat2022}'s thesis about the scales being created. Indeed, as explained, \emph{meaningfulness} is a form of invariance of the statements you make; it is not some ``quality'' of the numbers you assign and, certainly, it does not express how much sense those numbers may have for you.

Overall, we think that MA1 is not correct and does not hold since, for all the reasons explained above: even ``if you do know where the numbers came from'', this is not enough to them manipulate them as if ``numbers don't remember where they come from''. In other terms, you should still take into consideration the properties of the scale those numbers belong to and admissible operations on that scale.

\subsection{MA2: All IR effectiveness metrics can be considered to be interval scale measurements}
\label{subsec:MA2}

This argument by \citeauthor{Moffat2022} revolves around two severe misunderstandings. 

The first misconception is that an interval scale does not require equally spaced steps and, thus, whatever IR measure, even not equally spaced, can pretend to be an interval scale. In Section~\ref{subsec:scales}, we have explained how the \emph{solvability condition}, which express equi-spacing, is one of the axioms required to define an interval scale. Therefore, whatever numerical mapping not complying with the solvability condition does not match the definition and cannot be called an interval scale. On the other hand, a scale where order is preserved but intervals are not equi-spaced exists and it is an \emph{ordinal scale}, as also remarked by \citet[p.~679]{Stevens1946}: ``on an ordinal scale [...] the successive intervals on the scale are unequal in size''.

The second misconception is about what \emph{meaningfulness} is in the theory of measurement. \citeauthor{Moffat2022} just uses \texttt{meaningfulness} in its everyday language use of `making sense' or `being credible' along examples like when, in Sections II-D and II-E of his paper, the provost analyses the professor salaries or like when, in Section III-D of his paper, a market study leads to associate a \ac{SERP} with a numerical mapping corresponding to \ac{RR}. All of this has nothing to do with the notion of \emph{meaningfulness} in the theory of measurement and, as it should be clear from the discussion in the previous section, having some rationale in assigning numbers to objects does not imply or guarantee any scale properties, since much more precise conditions should be met in this case, such as the solvability condition. Therefore, a sentence like ``most current IR metrics are well-founded, and, moreover, [...] those metrics are more \texttt{meaningful} in their current form''~\citep[p.~105564]{Moffat2022} is wrong because \emph{meaningfulness} neither is a property of a measure nor it is a synonym of well-founded nor it something you can have ``more'' or ``less'', because either a statement is \emph{meaningful} (invariant) or it is not. Even more, it is not a transitive property you can use to justify some sort of chain of reasoning like: the assignment of numbers makes sense to me (this is not \emph{meaningfulness}), thus the measure makes sense to me (this is not \emph{meaningfulness}), thus the measure is an interval scale (this is neither \emph{meaningfulness} nor being an interval scale), thus the operations with that numbers make sense (this is not \emph{meaningfulness}), thus the statements/inferences make sense (here, in case, \emph{meaningfulness} would be about the invariance of the statement and not their sense or truth).

Overall, we think that MA2 is not correct and does not hold since, for all the reasons explained above: neither all \ac{IR} measures are interval scales nor can you always compute means over them and obtain \emph{meaningful} statements in a scientific sense nor interval scales with intervals not equi-spaced exist.

As a side note, for these reasons, the examples by~\citeauthor{Moffat2022} on professor salaries and \ac{SERP} pages/\ac{RR} are not interval scales, as a consequence means cannot be computed and the resulting statement cannot be \emph{meaningful}. 

\subsection{MA3: We believe that intervalization should be regarded with scepticism}
\label{subsec:MA3}

The intervalization procedure we proposed in our paper proceeds along this lines: (i) generate all the possible \ac{SERP}; (ii) compute the desired \ac{IR} evaluation measure; (iii) sort the \ac{SERP} according to the computed measure; (iv) use the rank assigned to a \ac{SERP} in (iii), which is an interval scale by construction, in the subsequent analyses and statistical test. Since, as already discussed in Section~\ref{subsec:MA2}, not all the IR evaluation measures can be considered to be interval scales, the proposed intervalization could find some useful application.

As we already discussed in Section~\ref{subsec:scales} the \emph{solvability condition}, i.e. equally spaced steps, is an axiom to be complied with for having an interval scale. Moreover, despite what \citeauthor{Moffat2022} reports about \citeauthor{Stevens1946}'s paper, \citeauthor{Stevens1946} actually says that equi-spacing is a requirement for an interval scale. Therefore, not holding, this should not be taken as a motivation for regarding the proposed intervalization with scepticism.  

In our paper, we do not claim that the purpose of intervalization is to make an IR evaluation measure a more ``defensible measurement of SERP usefulness''. We actually assume that every IR evaluation measure embeds its own user viewpoint -- being it defensible or not -- and, in case that a measure does not comply with the requirements for an interval scale, our intervalization procedure tries to preserve that user viewpoint as much as possible, still obtaining an interval scale, by keeping the same ordering of \ac{SERP} produced by that user viewpoint. Therefore, not holding, this should not be taken as a motivation for regarding the proposed intervalization with scepticism.  

The fact that the intervalization procedure changes the numerical mapping and that this will affect the subsequent computations is quite a trivial observation and it is exactly the purpose of this transformation. Indeed, our paper reports an extensive experimentation and a throughout investigation of the impacts and effects of this transformation on several kinds of statistical analyses and across several standard test collections. 

Overall, what our intervalization procedure gives you is the possibility of formulating \emph{meaningful} statements, doing its best at preserving the user viewpoint embedded by an \ac{IR} evaluation measure; it does not aim at all at making that user viewpoint more or less defensible. Therefore, researchers may or may not be interested in this approach and may or may not adopt it in their own experiments; researchers may hopefully also come up with other more brilliant solutions to ensure the \emph{meaningfulness} of our statements. Perhaps, researchers will not be sceptic about our intervalization procedure for the very same reasons suggested by~\citeauthor{Moffat2022} or, maybe, \citeauthor{Moffat2022} will provide some experimental evidence to substantiate his scepticism and, in case, to help identifying and quantifying specific issues and how to address them.

\subsection{Other Considerations}

\subsubsection*{Our line of work} 

There is some misunderstanding in~\citeauthor{Moffat2022}'s comment about how different lines of our work relate together, often mentioning them all together as if they were all the same, reporting the same claims, or as if all of them could be refuted by the same argument.

\citet{Fuhr2017} listed a series of practices he considers common mistakes in \ac{IR} evaluation, among which averaging \ac{RR} since it is not an interval scale, to be avoided. \citet{Sakai2020} already commented on the prescriptive nature of \citeauthor{Fuhr2017}'s paper and on what he considers or not to be mistakes; among them, \citeauthor{Sakai2020} considers averaging \ac{RR} a legitimate operation.

Our work commented by~\citeauthor{Moffat2022} actually originates from a different line of research. In our first work~\citep{FerranteEtAl2015b}, we started to seek for a way to apply the theory of measurement to the \ac{IR} evaluation by finding axioms (swap and replacement among relevant documents in a \ac{SERP}) that described the properties of a \ac{SERP} in the real word, a prerequisite for understanding the properties of \ac{IR} evaluation measures and their scales. Later on, in \citep{FerranteEtAl2017} and, more completely, in \citep{FerranteEtAl2018b} we used the notion of \emph{difference structure} explained in Section~\ref{subsec:scales} to formally describe the \ac{SERP} in the real world and from there to derive an interval scale measure; this, in turn, allowed for verifying which \ac{IR} evaluation measures were an interval scale (at least with respect to the identified difference structure) by seeking for an affine transformation. Then, in \citep{FerranteEtAl2019c} we started to explore the impact that using or not an interval scale can have in \ac{IR} experimentation. Finally, in our last work \citep{FerranteEtAl2021c} commented by~\citeauthor{Moffat2022}, we joined our interests and proposed \emph{meaningfulness} as a fundamental concept to be accounted for in \ac{IR} evaluation, we further investigated the implications of \ac{IR} measures being interval scales or not and why, and we proposed intervalization as a viable approach to ensure \emph{meaningfulness}. All in all, the objective of all these works is to provide better theoretical foundations to our evaluation methods and concrete means to achieve them. In this respect, these works are in the wake that others have also followed but using different approaches, such as \citet{vanRijsbergen1974}, \citet{Bollmann1984,BollmannCherniavsky1980}, or \citet{AmigoMizzaro2020}.

Finally, to the best of our knowledge, \citet{FerranteEtAl2019c,FerranteEtAl2021c} represent the first works to experimentally investigate in a systematic way the impact of scales and departure from their assumptions on different types of statistical analyses. 

In this perspective, the opening statement in the abstract of \citeauthor{Moffat2022}'s paper ``A sequence of recent papers, including in this journal, has considered the role of measurement scales in information retrieval (IR) experimentation, and presented the argument that (only) uniform-step interval scales should be used. Hence, it has been argued, well-known metrics such as reciprocal rank, expected reciprocal rank, normalized discounted cumulative gain, and average precision, should be either discarded as measurement tools, or adapted so that their metric values lie at uniformly-spaced points on the
number line'' is, at best, reductive of what our work actually is and shifts the focus of the discussion from laying theoretical and experimental foundations for our experimental methodology to ``use/do not use interval scales'' or ''use/do not use that evaluation measure''. 

\subsubsection*{Recall Base} 

When talking about the problems caused by the recall base (RB), i.e. the number of relevant documents for a topic, in Section IV-B of his paper, \citeauthor{Moffat2022} states that ``our contention in this work is that the measurement scale is always the positive real number line, and hence that no question of alignment (or not) of measurement points across sets of topics arises'' since, always according to \citeauthor{Moffat2022}, ``from the point of view of Fuhr and Ferrante et al., those difficulties arise because normalization by RB means that the set of generable measurement points for any query in a set of topics might not numerically align with the available measurement points for other topics that have different values for RB''. 

Actually, the view point expressed in our paper is that measures which explicitly depend on the recall base in their formulation lead to different scales on different topics and, as a consequence, they cannot be compared or mixed up, like you would not mix up length and mass, even if they are both ratio scales. So, the issue is more profound than a lack of numerical alignment.

Finally, in Section III-G of his paper, \citeauthor{Moffat2022} suggests that, differently from what reported in our previous works, also \ac{RBP} with $p = 0.5$ is not an interval scale when, for example, a \ac{SERP} is truncated at a length $k$ greater than the recall base for that topic. Actually, what our previous work shows is that \ac{RBP} with $p = 0.5$  is an interval scale when you construct it by assuming that as many relevant documents as needed are available, independently from the length of the \ac{SERP} or its truncation point. This is quite a reasonable assumption because you are creating a scale which should hold for all the topics and not a different \ac{RBP} scale for topics with one relevant document, topics with two relevant documents, and so on. It is the same line of reasoning you adopt when creating a scale for length: neither you create a scale for a geographical area with small trees and a separate one for another area with tall trees, nor you attempt to remove from a scale the ticks corresponding to some trees for discovering later on that then it is no more an interval scale. 

As a side note, from a more formal point of view, as discussed in Section~\ref{subsec:scales}, the \emph{solvability condition}~\cite[pp.~56ff.]{Rossi2014} and ~\citep[pp.~136ff.]{KrantzEtAl1971}, which leads to equally spaced steps, is one of the properties required for constructing an interval scale and, if your real world systems do not match it, e.g. because they are limited to the case of a single relevant document, it trivially follows that you can construct an interval scale. However, as explained before, this is not an issue of the numerical mapping but rather of the real world objects that lack the needed properties to create an interval scale.

\section{Concluding Remarks}
\label{sec:conclusions}

In this paper we have replied to~\citeauthor{Moffat2022}'s comment on our previous work~\cite{FerranteEtAl2021c}. In doing so, we have briefly summarised the main concepts of the representational theory of measurement~\citep{KrantzEtAl1971,LuceEtAl1990,SuppesEtAl1989} and of \emph{meaningfulness}~\cite{Narens2002} and we have explained why the main arguments by~\citeauthor{Moffat2022} do not hold, mainly because of misconceptions on these foundational concepts of the representational theory of measurment and of \emph{meaningfulness}.

As said, we really welcome~\citeauthor{Moffat2022}'s comment on our work because it offers the opportunity for an open discussion on important topics for our field. On the other hand, we remark that we would prefer avoiding to frame the discussion as the contrast between ``a bleak picture of past decades of IR evaluation'' and ``a more optimistic view of IR evaluation'', suggested by~\citeauthor{Moffat2022}. First, in our work we have never criticized or even deprecated past decades of IR evaluation. Saying that scale properties of evaluation measures matter or that \emph{meaningfulness} matters is neither criticizing nor deprecating past research, at least any more than talking about modern building techniques can be seen as a criticism of the pyramids. Nor we can consider as a criticism asking ourselves which statements in the \ac{IR} literature are \emph{meaningful}, i.e. invariant, because this, for example, would help in knowing what could potentially generalize in an easier way.  Second, framing the question as a contrast risks to amplify a ``defensive attitude'' in the field, rightly motivated by  safeguarding the seminal and paramount results of our past research, at the expense of an open-minded discussion of the topic and of a collaboration among researchers on how to develop and adopt better foundations for our evaluation methods.

A general feeling that emerges from \citeauthor{Moffat2022}'s comment is that we overlook the user viewpoint or we do not account for the user experience. On the contrary, we very much agree with \citeauthor{Moffat2022} and all the rest of the research community that the user satisfaction is the ultimate goal of our measurement and that evaluation measures should embed some user viewpoint. We just say that we should strive for this goal in the most sound and safe way possible; the proposed intervalization procedure is just a simple example of such an attempt. We actually do hope that this discussion and, especially, further research by others will deliver much better solutions in this respect.

Both here and in our work commented by~\citeauthor{Moffat2022}, we have indicated what this ``sound and safe way'' could be, i.e. a deeper investigation and adoption of the concept of \emph{meaningfulness}, at least in its scientific sense of \emph{invariance}. Indeed, \emph{meaningfulness} could be a way the achieve that \emph{generalizability} of results that we lack in our field. Moreover, \emph{meaningfulness} frees us from the debate on ``should we average or not?'' or ``should we give or adhere to prescriptions or not?'' and focus our attention on the real goal, i.e. \emph{drawing more robust and generalizable inferences and conclusions through better foundations of our evaluation methods}.

Finally, independently from the stance researchers may have on these topics, we think that more thorough experimentation should be carried out in the field also by others, in order to transfer theoretical models and considerations to practice, to gain a better understanding of the implications of the different choices, and to have a more informed discussion on the pros and cons of the various alternatives.

\acrodef{3G}[3G]{Third Generation Mobile System}
\acrodef{5S}[5S]{Streams, Structures, Spaces, Scenarios, Societies}
\acrodef{AA}[AA]{Active Agreements}
\acrodef{AAAI}[AAAI]{Association for the Advancement of Artificial Intelligence}
\acrodef{AAL}[AAL]{Annotation Abstraction Layer}
\acrodef{AAM}[AAM]{Automatic Annotation Manager}
\acrodef{AAP}[AAP]{Average Average Precision}
\acrodef{ACLIA}[ACLIA]{Advanced Cross-Lingual Information Access}
\acrodef{ACM}[ACM]{Association for Computing Machinery}
\acrodef{AD}[AD]{Active Disagreements}
\acrodef{ADSL}[ADSL]{Asymmetric Digital Subscriber Line}
\acrodef{ADUI}[ADUI]{ADministrator User Interface}
\acrodef{AI}[AI]{Artificial Intelligence}
\acrodef{AIP}[AIP]{Archival Information Package}
\acrodef{AJAX}[AJAX]{Asynchronous JavaScript Technology and \acs{XML}}
\acrodef{ALS}[ALS]{Amyotrophic Lateral Sclerosis}
\acrodef{ALSFRS-R}[ALSFRS-R]{ALS Functional Rating Scale Revisited}
\acrodef{ALU}[ALU]{Aritmetic-Logic Unit}
\acrodef{AMUSID}[AMUSID]{Adaptive MUSeological IDentity-service}
\acrodef{ANOVA}[ANOVA]{ANalysis Of VAriance}
\acrodef{ANSI}[ANSI]{American National Standards Institute}
\acrodef{AP}[AP]{Average Precision}
\acrodef{APC}[APC]{AP Correlation}
\acrodef{API}[API]{Application Program Interface}
\acrodef{AR}[AR]{Address Register}
\acrodef{AS}[AS]{Annotation Service}
\acrodef{ASAP}[ASAP]{Adaptable Software Architecture Performance}
\acrodef{ASI}[ASI]{Annotation Service Integrator}
\acrodef{ASL}[ASL]{Achieved Significance Level}
\acrodef{ASM}[ASM]{Annotation Storing Manager}
\acrodef{ASR}[ASR]{Automatic Speech Recognition}
\acrodef{ASUI}[ASUI]{ASsessor User Interface}
\acrodef{ATIM}[ATIM]{Annotation Textual Indexing Manager}
\acrodef{AUC}[AUC]{Area Under the ROC Curve}
\acrodef{AUI}[AUI]{Administrative User Interface}
\acrodef{AWARE}[AWARE]{Assessor-driven Weighted Averages for Retrieval Evaluation}
\acrodef{BANKS-I}[BANKS-I]{Browsing ANd Keyword Searching I}
\acrodef{BANKS-II}[BANKS-II]{Browsing ANd Keyword Searching II}
\acrodef{BH}[BH]{Benjamini-Hochberg}
\acrodef{bpref}[bpref]{Binary Preference}
\acrodef{BNF}[BNF]{Backus and Naur Form}
\acrodef{BPM}[BPM]{Bejeweled Player Model}
\acrodef{BRICKS}[BRICKS]{Building Resources for Integrated Cultural Knowledge Services}
\acrodef{CAN}[CAN]{Content Addressable Netword}
\acrodef{CAS}[CAS]{Content-And-Structure}
\acrodef{CBSD}[CBSD]{Component-Based Software Developlement}
\acrodef{CBSE}[CBSE]{Component-Based Software Engineering}
\acrodef{CB-SPE}[CB-SPE]{Component-Based \acs{SPE}}
\acrodef{CD}[CD]{Collaboration Diagram}
\acrodef{CD}[CD]{Compact Disk}
\acrodef{CDF}[CDF]{Cumulative Density Function}
\acrodef{CENL}[CENL]{Conference of European National Librarians}
\acrodef{CIDOC CRM}[CIDOC CRM]{CIDOC Conceptual Reference Model}
\acrodef{CIR}[CIR]{Current Instruction Register}
\acrodef{CIRCO}[CIRCO]{Coordinated Information Retrieval Components Orchestration}
\acrodef{CG}[CG]{Cumulated Gain}
\acrodef{CL}[CL]{Curriculum Learning}
\acrodef{CL-ESA}[CL-ESA]{Cross-Lingual Explicit Semantic Analysis}
\acrodef{CLAIRE}[CLAIRE]{Combinatorial visuaL Analytics system for Information Retrieval Evaluation}
\acrodef{CLEF1}[CLEF]{Cross-Language Evaluation Forum}
\acrodef{CLEF}[CLEF]{Conference and Labs of the Evaluation Forum}
\acrodef{CLIR}[CLIR]{Cross Language Information Retrieval}
\acrodef{CM}[CM]{Continuation Methods}
\acrodef{CMS}[CMS]{Content Management System}
\acrodef{CMT}[CMT]{Campaign Management Tool}
\acrodef{CNR}[CNR]{Italian National Council of Research}
\acrodef{CO}[CO]{Content-Only}
\acrodef{COD}[COD]{Code On Demand}
\acrodef{CODATA}[CODATA]{Committee on Data for Science and Technology}
\acrodef{COLLATE}[COLLATE]{Collaboratory for Annotation Indexing and Retrieval of Digitized Historical Archive Material}
\acrodef{CP}[CP]{Characteristic Pattern}
\acrodef{CPE}[CPE]{Control Processor Element}
\acrodef{CPU}[CPU]{Central Processing Unit}
\acrodef{CQL}[CQL]{Contextual Query Language}
\acrodef{CRP}[CRP]{Cumulated Relative Position}
\acrodef{CRUD}[CRUD]{Create--Read--Update--Delete}
\acrodef{CS}[CS]{Characteristic Structure}
\acrodef{CSM}[CSM]{Campaign Storing Manager}
\acrodef{CSS}[CSS]{Cascading Style Sheets}
\acrodef{CTR}[CTR]{Click-Through Rate}
\acrodef{CU}[CU]{Control Unit}
\acrodef{CUI}[CUI]{Client User Interface}
\acrodef{CV}[CV]{Cross-Validation}
\acrodef{DAFFODIL}[DAFFODIL]{Distributed Agents for User-Friendly Access of Digital Libraries}
\acrodef{DAO}[DAO]{Data Access Object}
\acrodef{DARE}[DARE]{Drawing Adequate REpresentations}
\acrodef{DARPA}[DARPA]{Defense Advanced Research Projects Agency}
\acrodef{DAS}[DAS]{Distributed Annotation System}
\acrodef{DB}[DB]{DataBase}
\acrodef{DBMS}[DBMS]{DataBase Management System}
\acrodef{DC}[DC]{Dublin Core}
\acrodef{DCG}[DCG]{Discounted Cumulated Gain}
\acrodef{DCMI}[DCMI]{Dublin Core Metadata Initiative}
\acrodef{DCV}[DCV]{Document Cut--off Value}
\acrodef{DD}[DD]{Deployment Diagram}
\acrodef{DDC}[DDC]{Dewey Decimal Classification}
\acrodef{DDS}[DDS]{Direct Data Structure}
\acrodef{DF}[DF]{Degrees of Freedom}
\acrodef{DFI}[DFI]{Divergence From Independence}
\acrodef{DFR}[DFR]{Divergence From Randomness}
\acrodef{DHT}[DHT]{Distributed Hash Table}
\acrodef{DI}[DI]{Digital Image}
\acrodef{DIKW}[DIKW]{Data, Information, Knowledge, Wisdom}
\acrodef{DIL}[DIL]{\acs{DIRECT} Integration Layer}
\acrodef{DiLAS}[DiLAS]{Digital Library Annotation Service}
\acrodef{DIRECT}[DIRECT]{Distributed Information Retrieval Evaluation Campaign Tool}
\acrodef{DKMS}[DKMS]{Data and Knowledge Management System}
\acrodef{DL}[DL]{Digital Library}
\acrodefplural{DL}[DL]{Digital Libraries}
\acrodef{DLMS}[DLMS]{Digital Library Management System}
\acrodef{DLOG}[DL]{Description Logics}
\acrodef{DLS}[DLS]{Digital Library System}
\acrodef{DLSS}[DLSS]{Digital Library Service System}
\acrodef{DM}[DM]{Data Mining}
\acrodef{DO}[DO]{Digital Object}
\acrodef{DOI}[DOI]{Digital Object Identifier}
\acrodef{DOM}[DOM]{Document Object Model}
\acrodef{DoMDL}[DoMDL]{Document Model for Digital Libraries}
\acrodef{DP}[DP]{Discriminative Power}
\acrodef{DPBF}[DPBF]{Dynamic Programming Best-First}
\acrodef{DR}[DR]{Data Register}
\acrodef{DRIVER}[DRIVER]{Digital Repository Infrastructure Vision for European Research}
\acrodef{DTD}[DTD]{Document Type Definition}
\acrodef{DVD}[DVD]{Digital Versatile Disk}
\acrodef{EAC-CPF}[EAC-CPF]{Encoded Archival Context for Corporate Bodies, Persons, and Families}
\acrodef{EAD}[EAD]{Encoded Archival Description}
\acrodef{EAN}[EAN]{International Article Number}
\acrodef{EBU}[EBU]{Expected Browsing Utility}
\acrodef{ECD}[ECD]{Enhanced Contenty Delivery}
\acrodef{ECDL}[ECDL]{European Conference on Research and Advanced Technology for Digital Libraries}
\acrodef{EDM}[EDM]{Europeana Data Model}
\acrodef{EG}[EG]{Execution Graph}
\acrodef{ELDA}[ELDA]{Evaluation and Language resources Distribution Agency}
\acrodef{ELRA}[ELRA]{European Language Resources Association}
\acrodef{EM}[EM]{Expectation Maximization}
\acrodef{EMMA}[EMMA]{Extensible MultiModal Annotation}
\acrodef{EPROM}[EPROM]{Erasable Programmable \acs{ROM}}
\acrodef{EQNM}[EQNM]{Extended Queueing Network Model}
\acrodef{ER}[ER]{Entity--Relationship}
\acrodef{ERR}[ERR]{Expected Reciprocal Rank}
\acrodef{ERS}[ERS]{Empirical Relational System}
\acrodef{ESA}[ESA]{Explicit Semantic Analysis}
\acrodef{ESL}[ESL]{Expected Search Length}
\acrodef{ETL}[ETL]{Extract-Transform-Load}
\acrodef{FAST}[FAST]{Flexible Annotation Service Tool}
\acrodef{FDR}[FDR]{False Discovery Rate}
\acrodef{FIFO}[FIFO]{First-In / First-Out}
\acrodef{FIRE}[FIRE]{Forum for Information Retrieval Evaluation}
\acrodef{FN}[FN]{False Negative}
\acrodef{FNR}[FNR]{False Negative Rate}
\acrodef{FOAF}[FOAF]{Friend of a Friend}
\acrodef{FORESEE}[FORESEE]{FOod REcommentation sErvER}
\acrodef{FP}[FP]{False Positive}
\acrodef{FPR}[FPR]{False Positive Rate}
\acrodef{FVC}[FVC]{Forced Vital Capacity}
\acrodef{FWER}[FWER]{Family-wise Error Rate}
\acrodef{GIF}[GIF]{Graphics Interchange Format}
\acrodef{GIR}[GIR]{Geografic Information Retrieval}
\acrodef{GAP}[GAP]{Graded Average Precision}
\acrodef{GLM}[GLM]{General Linear Model}
\acrodef{GLMM}[GLMM]{General Linear Mixed Model}
\acrodef{GMAP}[GMAP]{Geometric Mean Average Precision}
\acrodef{GoP}[GoP]{Grid of Points}
\acrodef{GPRS}[GPRS]{General Packet Radio Service}
\acrodef{gP}[gP]{Generalized Precision}
\acrodef{gR}[gR]{Generalized Recall}
\acrodef{gRBP}[gRBP]{Graded Rank-Biased Precision}
\acrodef{GT}[GT]{Generalizability Theory}
\acrodef{GTIN}[GTIN]{Global Trade Item Number}
\acrodef{GUI}[GUI]{Graphical User Interface}
\acrodef{GW}[GW]{Gateway}
\acrodef{HCI}[HCI]{Human Computer Interaction}
\acrodef{HDS}[HDS]{Hybrid Data Structure}
\acrodef{HIR}[HIR]{Hypertext Information Retrieval}
\acrodef{HIT}[HIT]{Human Intelligent Task}
\acrodef{HITS}[HITS]{Hyperlink-Induced Topic Search}
\acrodef{HMM}[HMM]{Hidden Markov Model}
\acrodef{HTML}[HTML]{HyperText Markup Language}
\acrodef{HTTP}[HTTP]{HyperText Transfer Protocol}
\acrodef{HSD}[HSD]{Honestly Significant Difference}
\acrodef{ICA}[ICA]{International Council on Archives}
\acrodef{ICSU}[ICSU]{International Council for Science}
\acrodef{IDF}[IDF]{Inverse Document Frequency}
\acrodef{iDPP}[iDPP@CLEF]{Intelligent Disease Progression Prediction at CLEF}
\acrodef{IDS}[IDS]{Inverse Data Structure}
\acrodef{IEEE}[IEEE]{Institute of Electrical and Electronics Engineers}
\acrodef{IEI}[IEI]{Istituto della Enciclopedia Italiana fondata da Giovanni Treccani}
\acrodef{IETF}[IETF]{Internet Engineering Task Force}
\acrodef{IIR}[IIR]{Interactive Information Retrieval}
\acrodef{IMS}[IMS]{Information Management System}
\acrodef{IMSPD}[IMS]{Information Management Systems Research Group}
\acrodef{indAP}[indAP]{Induced Average Precision}
\acrodef{infAP}[infAP]{Inferred Average Precision}
\acrodef{INEX}[INEX]{INitiative for the Evaluation of \acs{XML} Retrieval}
\acrodef{INS-M}[INS-M]{Inverse Set Data Model}
\acrodef{INTR}[INTR]{Interrupt Register}
\acrodef{IP}[IP]{Internet Protocol}
\acrodef{IPSA}[IPSA]{Imaginum Patavinae Scientiae Archivum}
\acrodef{IR}[IR]{Information Retrieval}
\acrodef{IRON}[IRON]{Information Retrieval ON}
\acrodef{IRON2}[IRON$^2$]{Information Retrieval On aNNotations}
\acrodef{IRON-SAT}[IRON-SAT]{\acs{IRON} - Statistical Analysis Tool}
\acrodef{IRS}[IRS]{Information Retrieval System}
\acrodef{ISAD(G)}[ISAD(G)]{International Standard for Archival Description (General)}
\acrodef{ISBN}[ISBN]{International Standard Book Number}
\acrodef{ISIS}[ISIS]{Interactive SImilarity Search}
\acrodef{ISJ}[ISJ]{Interactive Searching and Judging}
\acrodef{ISO}[ISO]{International Organization for Standardization}
\acrodef{ITU}[ITU]{International Telecommunication Union }
\acrodef{ITU-T}[ITU-T]{Telecommunication Standardization Sector of \acs{ITU}}
\acrodef{IV}[IV]{Information Visualization}
\acrodef{JAN}[JAN]{Japanese Article Number}
\acrodef{JDBC}[JDBC]{Java DataBase Connectivity}
\acrodef{JMB}[JMB]{Java--Matlab Bridge}
\acrodef{JPEG}[JPEG]{Joint Photographic Experts Group}
\acrodef{JSON}[JSON]{JavaScript Object Notation}
\acrodef{JSP}[JSP]{Java Server Pages}
\acrodef{JTE}[JTE]{Java-Treceval Engine}
\acrodef{KDE}[KDE]{Kernel Density Estimation}
\acrodef{KLD}[KLD]{Kullback-Leibler Divergence}
\acrodef{KLAPER}[KLAPER]{Kernel LAnguage for PErformance and Reliability analysis}
\acrodef{LAM}[LAM]{Libraries, Archives, and Museums}
\acrodef{LAM2}[LAM]{Logistic Average Misclassification}
\acrodef{LAN}[LAN]{Local Area Network}
\acrodef{LD}[LD]{Linked Data}
\acrodef{LEAF}[LEAF]{Linking and Exploring Authority Files}
\acrodef{LIDO}[LIDO]{Lightweight Information Describing Objects}
\acrodef{LIFO}[LIFO]{Last-In / First-Out}
\acrodef{LM}[LM]{Language Model}
\acrodef{LMT}[LMT]{Log Management Tool}
\acrodef{LOD}[LOD]{Linked Open Data}
\acrodef{LODE}[LODE]{Linking Open Descriptions of Events}
\acrodef{LpO}[LpO]{Leave-$p$-Out}
\acrodef{LRM}[LRM]{Local Relational Model}
\acrodef{LRU}[LRU]{Last Recently Used}
\acrodef{LS}[LS]{Lexical Signature}
\acrodef{LSM}[LSM]{Log Storing Manager}
\acrodef{LtR}[LtR]{Learning to Rank}
\acrodef{LUG}[LUG]{Lexical Unit Generator}
\acrodef{MA}[MA]{Mobile Agent}
\acrodef{MA}[MA]{Moving Average}
\acrodef{MACS}[MACS]{Multilingual ACcess to Subjects}
\acrodef{MADCOW}[MADCOW]{Multimedia Annotation of Digital Content Over the Web}
\acrodef{MAD}[MAD]{Mean Assessed Documents}
\acrodef{MADP}[MADP]{Mean Assessed Documents Precision}
\acrodef{MADS}[MADS]{Metadata Authority Description Standard}
\acrodef{MAP}[MAP]{Mean Average Precision}
\acrodef{MARC}[MARC]{Machine Readable Cataloging}
\acrodef{MATTERS}[MATTERS]{MATlab Toolkit for Evaluation of information Retrieval Systems}
\acrodef{MDA}[MDA]{Model Driven Architecture}
\acrodef{MDD}[MDD]{Model-Driven Development}
\acrodef{METS}[METS]{Metadata Encoding and Transmission Standard}
\acrodef{MIDI}[MIDI]{Musical Instrument Digital Interface}
\acrodef{MIME}[MIME]{Multipurpose Internet Mail Extensions}
\acrodef{ML}[ML]{Machine Learning}
\acrodef{MLE}[MLE]{Maximum Likelihood Estimation}
\acrodef{MLIA}[MLIA]{MultiLingual Information Access}
\acrodef{MM}[MM]{Machinery Model}
\acrodef{MMU}[MMU]{Memory Management Unit}
\acrodef{MODS}[MODS]{Metadata Object Description Schema}
\acrodef{MOF}[MOF]{Meta-Object Facility}
\acrodef{MP}[MP]{Markov Precision}
\acrodef{MPEG}[MPEG]{Motion Picture Experts Group}
\acrodef{MRD}[MRD]{Machine Readable Dictionary}
\acrodef{MRF}[MRF]{Markov Random Field}
\acrodef{MRR}[MRR]{Mean Reciprocal Rank}
\acrodef{MS}[MS]{Mean Squares}
\acrodef{MS2}[MS]{Multiple Sclerosis}
\acrodef{MSAC}[MSAC]{Multilingual Subject Access to Catalogues}
\acrodef{MSE}[MSE]{Mean Square Error}
\acrodef{MT}[MT]{Machine Translation}
\acrodef{MV}[MV]{Majority Vote}
\acrodef{MVC}[MVC]{Model-View-Controller}
\acrodef{NACSIS}[NACSIS]{NAtional Center for Science Information Systems}
\acrodef{NAP}[NAP]{Network processors Applications Profile}
\acrodef{NCP}[NCP]{Normalized Cumulative Precision}
\acrodef{nCG}[nCG]{Normalized Cumulated Gain}
\acrodef{nCRP}[nCRP]{Normalized Cumulated Relative Position}
\acrodef{nDCG}[nDCG]{Normalized Discounted Cumulated Gain}
\acrodef{nMCG}[nMCG]{Normalized Markov Cumulated Gain}
\acrodef{NESTOR}[NESTOR]{NEsted SeTs for Object hieRarchies}
\acrodef{NEXI}[NEXI]{Narrowed Extended XPath I}
\acrodef{NII}[NII]{National Institute of Informatics}
\acrodef{NISO}[NISO]{National Information Standards Organization}
\acrodef{NIST}[NIST]{National Institute of Standards and Technology}
\acrodef{NIV}[NIV]{Non-Invasive Ventilation}
\acrodef{NLP}[NLP]{Natural Language Processing}
\acrodef{NN}[NN]{Neural Network}
\acrodef{NP}[NP]{Network Processor}
\acrodef{NR}[NR]{Normalized Recall}
\acrodef{NRS}[NRS]{Numerical Relational System}
\acrodef{NS-M}[NS-M]{Nested Set Model}
\acrodef{NTCIR}[NTCIR]{NII Testbeds and Community for Information access Research}
\acrodef{OAI}[OAI]{Open Archives Initiative}
\acrodef{OAI-ORE}[OAI-ORE]{Open Archives Initiative Object Reuse and Exchange}
\acrodef{OAI-PMH}[OAI-PMH]{Open Archives Initiative Protocol for Metadata Harvesting}
\acrodef{OAIS}[OAIS]{Open Archival Information System}
\acrodef{OC}[OC]{Operation Code}
\acrodef{OCLC}[OCLC]{Online Computer Library Center}
\acrodef{OMG}[OMG]{Object Management Group}
\acrodef{OO}[OO]{Object Oriented}
\acrodef{OODB}[OODB]{Object-Oriented \acs{DB}}
\acrodef{OODBMS}[OODBMS]{Object-Oriented \acs{DBMS}}
\acrodef{OPAC}[OPAC]{Online Public Access Catalog}
\acrodef{OQL}[OQL]{Object Query Language}
\acrodef{ORP}[ORP]{Open Relevance Project}
\acrodef{OSIRIS}[OSIRIS]{Open Service Infrastructure for Reliable and Integrated process Support}
\acrodef{P}[P]{Precision}
\acrodef{P2P}[P2P]{Peer-To-Peer}
\acrodef{PA}[PA]{Passive Agreements}
\acrodef{PAMT}[PAMT]{Pool-Assessment Management Tool}
\acrodef{PASM}[PASM]{Pool-Assessment Storing Manager}
\acrodef{PC}[PC]{Program Counter}
\acrodef{PCP}[PCP]{Pre-Commercial Procurement}
\acrodef{PCR}[PCR]{Peripherical Command Register}
\acrodef{PD}[PD]{Passive Disagreements}
\acrodef{PDA}[PDA]{Personal Digital Assistant}
\acrodef{PDF}[PDF]{Probability Density Function}
\acrodef{PDR}[PDR]{Peripherical Data Register}
\acrodef{PEG}[PEG]{Percutaneous Endoscopic Gastrostomy}
\acrodef{PIR}[PIR]{Personalized Information Retrieval}
\acrodef{POI}[POI]{\acs{PURL}-based Object Identifier}
\acrodef{PoS}[PoS]{Part of Speech}
\acrodef{PAA}[PAA]{Proportion of Active Agreements}
\acrodef{PPA}[PPA]{Proportion of Passive Agreements}
\acrodef{PPE}[PPE]{Programmable Processing Engine}
\acrodef{PREFORMA}[PREFORMA]{PREservation FORMAts for culture information/e-archives}
\acrodef{PRIMAD}[PRIMAD]{Platform, Research goal, Implementation, Method, Actor, and Data}
\acrodef{PRIMAmob-UML}[PRIMAmob-UML]{mobile \acs{PRIMA-UML}}
\acrodef{PRIMA-UML}[PRIMA-UML]{PeRformance IncreMental vAlidation in \acs{UML}}
\acrodef{PROM}[PROM]{Programmable \acs{ROM}}
\acrodef{PROMISE}[PROMISE]{Participative Research labOratory  for Multimedia and Multilingual Information Systems Evaluation}
\acrodef{pSQL}[pSQL]{propagate \acs{SQL}}
\acrodef{PUI}[PUI]{Participant User Interface}
\acrodef{PURL}[PURL]{Persistent \acs{URL}}
\acrodef{QA}[QA]{Question Answering}
\acrodef{QE}[QE]{Query Expansion}
\acrodef{QoS-UML}[QoS-UML]{\acs{UML} Profile for QoS and Fault Tolerance}
\acrodef{QPA}[QPA]{Query Performance Analyzer}
\acrodef{QPP}[QPP]{Query Performance Prediction}
\acrodef{R}[R]{Recall}
\acrodef{RAM}[RAM]{Random Access Memory}
\acrodef{RAMM}[RAM]{Random Access Machine}
\acrodef{RBO}[RBO]{Rank-Biased Overlap}
\acrodef{RBP}[RBP]{Rank-Biased Precision}
\acrodef{RBTO}[RBTO]{Rank-Based Total Order}
\acrodef{RDBMS}[RDBMS]{Relational \acs{DBMS}}
\acrodef{RDF}[RDF]{Resource Description Framework}
\acrodef{REST}[REST]{REpresentational State Transfer}
\acrodef{REV}[REV]{Remote Evaluation}
\acrodef{RF}[RF]{Relevance Feedback}
\acrodef{RFC}[RFC]{Request for Comments}
\acrodef{RIA}[RIA]{Reliable Information Access}
\acrodef{RMSE}[RMSE]{Root Mean Square Error}
\acrodef{RMT}[RMT]{Run Management Tool}
\acrodef{ROM}[ROM]{Read Only Memory}
\acrodef{ROMIP}[ROMIP]{Russian Information Retrieval Evaluation Seminar}
\acrodef{RoMP}[RoMP]{Rankings of Measure Pairs}
\acrodef{RoS}[RoS]{Rankings of Systems}
\acrodef{RP}[RP]{Relative Position}
\acrodef{RR}[RR]{Reciprocal Rank}
\acrodef{RSM}[RSM]{Run Storing Manager}
\acrodef{RST}[RST]{Rhetorical Structure Theory}
\acrodef{RSV}[RSV]{Retrieval Status Value}
\acrodef{RT-UML}[RT-UML]{\acs{UML} Profile for Schedulability, Performance and Time}
\acrodef{SA}[SA]{Software Architecture}
\acrodef{SAL}[SAL]{Storing Abstraction Layer}
\acrodef{SAMT}[SAMT]{Statistical Analysis Management Tool}
\acrodef{SAN}[SAN]{Sistema Archivistico Nazionale}
\acrodef{SASM}[SASM]{Statistical Analysis Storing Manager}
\acrodef{SBTO}[SBTO]{Set-Based Total Order}
\acrodef{SD}[SD]{Sequence Diagram}
\acrodef{SE}[SE]{Search Engine}
\acrodef{SEBD}[SEBD]{Convegno Nazionale su Sistemi Evoluti per Basi di Dati}
\acrodef{SEM}[SEM]{Standard Error of the Mean}
\acrodef{SERP}[SERP]{Search Engine Result Page}
\acrodef{SFT}[SFT]{Satisfaction--Frustration--Total}
\acrodef{SIL}[SIL]{Service Integration Layer}
\acrodef{SIP}[SIP]{Submission Information Package}
\acrodef{SKOS}[SKOS]{Simple Knowledge Organization System}
\acrodef{SM}[SM]{Software Model}
\acrodef{SME}[SME]{Statistics--Metrics-Experiments}
\acrodef{SMART}[SMART]{System for the Mechanical Analysis and Retrieval of Text}
\acrodef{SoA}[SoA]{Service-oriented Architectures}
\acrodef{SOA}[SOA]{Strength of Association}
\acrodef{SOAP}[SOAP]{Simple Object Access Protocol}
\acrodef{SOM}[SOM]{Self-Organizing Map}
\acrodef{SPARQL}[SPARQL]{Simple Protocol and RDF Query Language}
\acrodef{SPE}[SPE]{Software Performance Engineering}
\acrodef{SPINA}[SPINA]{Superimposed Peer Infrastructure for iNformation Access}
\acrodef{SPLIT}[SPLIT]{Stemming Program for Language Independent Tasks}
\acrodef{SPOOL}[SPOOL]{Simultaneous Peripheral Operations On Line}
\acrodef{SQL}[SQL]{Structured Query Language}
\acrodef{SR}[SR]{Sliding Ratio}
\acrodef{sRBP}[sRBP]{Session Rank Biased Precision}
\acrodef{SRU}[SRU]{Search/Retrieve via \acs{URL}}
\acrodef{SS}[SS]{Sum of Squares}
\acrodef{SSD}[s.s.d.]{statistically significantly different}
\acrodef{SSTF}[SSTF]{Shortest Seek Time First}
\acrodef{STAR}[STAR]{Steiner-Tree Approximation in Relationship graphs}
\acrodef{STON}[STON]{STemming ON}
\acrodef{SVM}[SVM]{Support Vector Machine}
\acrodef{TAC}[TAC]{Text Analysis Conference}
\acrodef{TBG}[TBG]{Time-Biased Gain}
\acrodef{TCP}[TCP]{Transmission Control Protocol}
\acrodef{TEL}[TEL]{The European Library}
\acrodef{TERRIER}[TERRIER]{TERabyte RetrIEveR}
\acrodef{TF}[TF]{Term Frequency}
\acrodef{TFR}[TFR]{True False Rate}
\acrodef{TLD}[TLD]{Top Level Domain}
\acrodef{TME}[TME]{Topics--Metrics-Experiments}
\acrodef{TN}[TN]{True Negative}
\acrodef{TO}[TO]{Transfer Object}
\acrodef{TP}[TP]{True Positve}
\acrodef{TPR}[TPR]{True Positive Rate}
\acrodef{TRAT}[TRAT]{Text Relevance Assessing Task}
\acrodef{TREC}[TREC]{Text REtrieval Conference}
\acrodef{TRECVID}[TRECVID]{TREC Video Retrieval Evaluation}
\acrodef{TTL}[TTL]{Time-To-Live}
\acrodef{UCD}[UCD]{Use Case Diagram}
\acrodef{UDC}[UDC]{Universal Decimal Classification}
\acrodef{uGAP}[uGAP]{User-oriented Graded Average Precision}
\acrodef{UI}[UI]{User Interface}
\acrodef{UML}[UML]{Unified Modeling Language}
\acrodef{UMT}[UMT]{User Management Tool}
\acrodef{UMTS}[UMTS]{Universal Mobile Telecommunication System}
\acrodef{UoM}[UoM]{Utility-oriented Measurement}
\acrodef{UPC}[UPC]{Universal Product Code}
\acrodef{URI}[URI]{Uniform Resource Identifier}
\acrodef{URL}[URL]{Uniform Resource Locator}
\acrodef{URN}[URN]{Uniform Resource Name}
\acrodef{USM}[USM]{User Storing Manager}
\acrodef{VA}[VA]{Visual Analytics}
\acrodef{VAIRE}[VAIR\"{E}]{Visual Analytics for Information Retrieval Evaluation}
\acrodef{VATE}[VATE$^2$]{Visual Analytics Tool for Experimental Evaluation}
\acrodef{VIRTUE}[VIRTUE]{Visual Information Retrieval Tool for Upfront Evaluation}
\acrodef{VD}[VD]{Virtual Document}
\acrodef{VDM}[VDM]{Visual Data Mining}
\acrodef{VIAF}[VIAF]{Virtual International Authority File}
\acrodef{VIM}[VIM]{International Vocabulary of Metrology}
\acrodef{VL}[VL]{Visual Language}
\acrodef{VoIP}[VoIP]{Voice over IP}
\acrodef{VS}[VS]{Visual Sentence}
\acrodef{W3C}[W3C]{World Wide Web Consortium}
\acrodef{WAN}[WAN]{Wide Area Network}
\acrodef{WHO}[WHO]{World Health Organization}
\acrodef{WLAN}[WLAN]{Wireless \acs{LAN}}
\acrodef{WP}[WP]{Work Package}
\acrodef{WS}[WS]{Web Services}
\acrodef{WSD}[WSD]{Word Sense Disambiguation}
\acrodef{WSDL}[WSDL]{Web Services Description Language}
\acrodef{WWW}[WWW]{World Wide Web}
\acrodef{XAI}[XAI]{eXplainable \acs{AI}}
\acrodef{XMI}[XMI]{\acs{XML} Metadata Interchange}
\acrodef{XML}[XML]{eXtensible Markup Language}
\acrodef{XPath}[XPath]{XML Path Language}
\acrodef{XSL}[XSL]{eXtensible Stylesheet Language}
\acrodef{XSL-FO}[XSL-FO]{\acs{XSL} Formatting Objects}
\acrodef{XSLT}[XSLT]{\acs{XSL} Transformations}
\acrodef{YAGO}[YAGO]{Yet Another Great Ontology}
\acrodef{YASS}[YASS]{Yet Another Suffix Stripper}


\end{document}